# Implementation of Formal Standard for Interoperability in M&S/Systems of Systems Integration with DEVS/SOA

*Saurabh Mittal*

Dunip Technologies
New Delhi, India
saurabh.mittal@duniptechnologies.com

*Bernard P. Zeigler*

Arizona Center for Integrative Modeling and Simulation,
University of Arizona,
Tucson, AZ, USA
zeigler@ece.arizona.edu

*Jose L. Risco-Martin*

Departamento de Arquitectura de Computadores y Automática
Facultad de Informática
Universidad Complutense de Madrid
Madrid, Spain
jlrisco@gmail.com

## Abstract
*Modeling and Simulation (M&S) is finding increasing application in development and testing of command and control systems comprised of information-intensive component systems. Achieving interoperability is one of the chief System of systems (SoS) engineering objectives in the development of command and control (C2) capabilities for joint and coalition warfare. In this paper, we apply an SoS perspective on the integration of M&S with such systems. We employ recently developed interoperability concepts based on linguistic categories along with the Discrete Event System Specification (DEVS) formalism to implement a standard for interoperability. We will show how the developed standard is implemented in DEVS/SOA net-centric modeling and simulation framework that uses XML-based Service Oriented Architecture (SOA). We will discuss the simulator interfaces and the design issues in their implementation in DEVS/SOA. We will illustrate the application of DEVS/SOA in a multi-agent test instrumentation system that is deployable as a SOA.*

## 1. Introduction
Modeling and Simulation (M&S) is finding increasing application in important aspects of command and control systems comprised of information intensive component systems. One aspect of such application is the incorporation of M&S functionality into such systems which is also an objective of the Extensible Modeling and Simulation Framework (XMSF[1]). Another aspect is the use of M&S to support the development and testing such systems as instances of System of Systems (SoS). The SoS concept relates to the attempt to integrate disparate systems to achieve a specific goal, typically not co-incident with the goals of the pre-existing component systems. Consequently, the defining concern in SoS engineering is interoperability, or lack thereof, among the constituent system [1, 2]. Achieving such interoperability is among the chief SoS engineering objectives in the development of command and control (C2) capabilities for joint and coalition warfare [3]. Sage [1] analogized the construction of SoS to the federation of socio-political systems and drew a parallel between such processes and the federation that is supported by the High Level Architecture (HLA, an IEEE standard fostered by the DoD to enable interoperation of simulation components [4]). In this light, the present author discussed the role that M&S can play in helping to address the interoperability problems in SoS engineering [5]. The present paper builds upon this work by considering not only the parallel between SoS engineering and distributed simulation, but also how M&S can be more integrally included within SoS engineering approaches. The focus of this paper is to present fundamental concepts to help tackle the integration of M&S and C2 SoS through the use of concepts and standards for interoperability based on the Discrete Event Systems Specification (DEVS) formalism. Our ultimate motivation is to apply M&S concepts and technologies to support collaborative decision making in C2 SoS as well as the testing and evaluation of such systems.

---
[1] XMSF: A set of Web-based technologies and distributed testbed [1]

DEVS environments such as DEVSJAVA, DEVS-C++, and others [28] are embedded in object-oriented implementations; they support the goal of representing executable model architectures in an object-oriented representational language. As a mathematical formalism, DEVS is platform independent, and its implementations adhere to the DEVS protocol so that DEVS models easily translate from one form (e.g., C++) to another (e.g., Java) [16]. Moreover, DEVS environments, such as DEVSJAVA, execute on commercial, off-the-shelf desktops or workstations and employ state-of-the-art libraries to produce graphical output.. DEVS environments are typically open architectures that have been extended to execute on various middleware such as the DoD's HLA standard, CORBA, SOAP, and others and can be readily interfaced to other engineering and simulation and modeling tools [29,30,31,32]. Furthermore, DEVS operation over web middleware (SOAP) enables it to utilize the net-centric environment of the Global Information Grid/Service Oriented Architecture (GIG/SOA). As a result of recent advances, DEVS can support model continuity through a simulation-based development and testing life cycle [29]. This means that the mapping of high-level requirement specifications into lower-level DEVS formalizations enables such specifications to be thoroughly tested in virtual simulation environments before being easily and consistently transitioned to operate in a real environment for further testing and fielding.

This article is an extension of a recent article by authors where the Standard for DEVS M&S interoperability was proposed [54]. The present work is a realization of the concepts in [54].

## 2. Interoperability in Distributed Simulation

As illustrated in Figure 1, HLA is a network middleware layer that supports message exchanges among simulation components, called federates, in a neutral format and also provides a range of services to support dynamic and efficient execution of simulations. However, experience with HLA has been disappointing and forced proponents to acknowledge the difference between enabling heterogeneous simulations to exchange data, so-called *technical* interoperability, and *substantive* interoperability – the desired outcome of exchanging meaningful data so that coherent interaction among federates takes place [5]. Tolk introduced the Levels of Conceptual Interoperability Model (LCIM) which identified seven levels of interoperability among participating systems [6]. These levels also can be viewed as a refinement of the operational interoperability type which is one of three defined by Dimario [7]. The *operational* type concerns linkages between systems in their interactions with one another, the environment, and with users. The additional levels provide more elaboration to the catch-all category of substantive interoperability and, as illustrated in Figure 1, are missing from HLA standard as such.

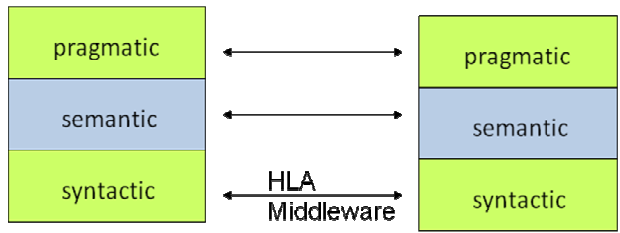

Figure 1. HLA Technical Interoperability

## 3. Levels of Conceptual Interoperability Model

Although Levels of Information Systems Interoperability [8] models are used successfully to determine the degree of interoperability between information technology systems, they do not provide a systematic formulation of the underlying properties of information exchange. To remedy this situation, the LCIM outlined in Table 1, was developed to become a bridge between conceptual and technical design for implementation, integration, or federation [9, 10].

The last column lists key conditions that are required to reach an interoperability level from the one below. Of course, the conditions accumulate as the level increases. We note that the conditions given in the LCIM for pragmatic interoperability require that the use of data be mutually understood, where the term "use" is interpreted as the context of its application. A reformulation of LCIM was presented in [11] where more definitive concepts for pragmatic interoperability including the concepts of pragmatic frames and pragmatic equivalence. Moreover, the definition of the semantic level requires the use of a single reference semantic model as a hub for information exchange among participants in collaboration. However such a hub and spokes approach, while desirable, is not always feasible. [12] evaluated a common information exchange model, C2IEDM, as an interoperability-enabling ontology for command and control. The conclusion is that even if there is room for improvements, the model supports almost all basic needs for such a semantic bridge. However, [13] claim that in its current form, the model is unbalanced in its levels of detail and too large to be practical. In the stratification to be introduced below, we review a more streamlined and extended account of information exchange levels.

| Level of Conceptual Interoperability | Characteristic | Key Condition |
|---|---|---|
| Conceptual | The assumptions and constraints underlying the meaningful abstraction of reality are aligned | Requires that conceptual models be documented based on engineering methods enabling their interpretation and evaluation by other engineers. |
| Dynamic | Participants are able to comprehend changes in system state and assumptions and constraints that each is making over time, and are able to take advantage of those changes. | Requires common understanding of system dynamics |
| Pragmatic | Participants are aware of the methods and procedures that each is employing | Requires that the use of the data – or the context of their application – is understood by the participating systems. |
| Semantic | The meaning of the data is shared | Requires a common information exchange reference model |
| Syntactic | Introduces a common structure to exchange information, | Requires that a common data format is used |
| Technical | Data can be exchanged between participants | Requires that a communication protocol exists |
| Stand alone | No interoperability | |

**Table 1**: Levels of Conceptual Interoperability

## 4. Linguistic Levels

The definitions given in [11] agree in general, but differ substantially, with those used in the LCIM. They are summarized:

- *Pragmatics*: Data use in relation to data structure and context of application
- *Semantics*: Low level semantics focuses on definitions and attributes of terms; high level semantics focuses on the combined meaning of multiple terms (Generalized Context). Note in contrast to the LCIM requirement for semantic interoperability, this definition focuses on the underlying requirement for achieving shared meanings rather than how this requirement is achieved.
- *Syntax* focuses on a structure and adherence to the rules that govern that structure, e.g., XML (Rules and Structure)

The authors of LCIM associate the lower layers with the problems of simulation interoperation while the upper layers relate to the problems of reuse and composition of models [14,15]. They conclude "simulation systems are based on models and their assumptions and constraints. If two simulation systems are combined, these assumptions and constraints must be aligned accordingly to ensure meaningful results."[10]. This suggests that levels of interoperability that have been identified in the area of M&S can serve as guidelines to discussion of information exchange in general. Therefore, we consider an earlier developed conceptual layered architecture for M&S [16]. We'll correlate the above linguistic definitions with the layers outlined below and shown in Figure 2.

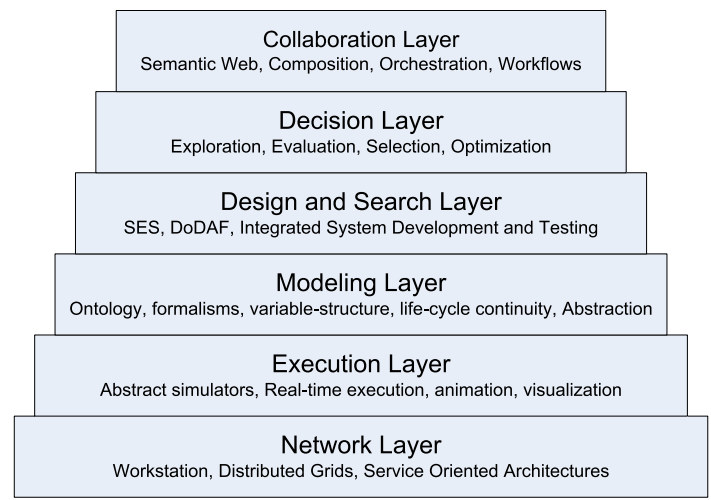

Figure 2. Architecture for Modeling and Simulation

*Network Layer* contains the actual computers (including workstations and high performance systems) and the connecting networks (both LAN and WAN, their hardware and software) that do the work of supporting all aspects of the M&S lifecycle.

*Execution Layer* is the software that executes the models in simulation time and/or real time to generate their behavior. Included in this layer are the protocols that provide the basis for distributed simulation (such as those that are standardized in the HLA. Also included are database management systems, software systems to support control of simulation executions, visualization and animation of the generated behaviors.

*Modeling Layer* supports the development of models in formalisms that are independent of any given simulation layer implementation. HLA just mentioned also provides object-oriented templates for model description aimed at supporting confederations of globally dispersed models. However, beyond this, the formalisms for model behavior, whether continuous, discrete or discrete event in nature) as well as structure change, are also included in this layer. Model construction and especially, the key processes of model abstraction and continuity over the lifecycle are also included. We also add ontologies to this layer where they are understood as models of the world for a particular conceptualization intended to support information exchange.

*Design and Search Layer* supports the design of systems, such as in the Department of Defense Architecture Framework (DoDAF) version 1.5 [55] where the design is based on specifying desired behaviors through models and implementing these behaviors through interconnection of system components. It also includes investigation of large families of alternative models, whether in the form of spaces set up by parameters or more powerful means of specifying alternative model structures such as provided by the SES methodology [11]. Artificial intelligence and simulated natural intelligence (evolutionary programming) may be brought in to help deal with combinatorial explosions occasioned by powerful model synthesizing capabilities.

*Decision Layer* applies the capability to search and simulate large model sets at the layer below to make decisions in solving real-world problems. Included are course-of-action planning, selection of design alternatives and other choices where the outcomes may be supported by concept explorations, "what-if" investigations, and optimizations of the models constructed in the modeling layer using the simulation layer below it.

*Collaboration Layer* enables people or intelligent agents with partial knowledge about a system, whether based on discipline, location, task, or responsibility specialization, to bring to bear individual perspectives and contributions to achieve an overall goal.

Using the definitions for linguistic levels above, we correlate such levels with the layers just discussed. As illustrated in Figure 3, at the syntactic level we associate network and execution layers. The semantic level corresponds with the modeling layer – where we have included ontology frameworks as well as dynamic system formalisms as models. Finally, the pragmatic level includes use of the information such as identified in the upper layers of the M&S architecture. This use occurs for example, in design and search, making decisions and collaborating to achieve common goals. Indeed, such mental activities, along with real-world physical actions that they lead to, provide the basis for enumerating the kinds of pragmatic frames that might be of interest in particular applications – the context of use.

The resulting stratification leads us to propose Table 2 for defining effective interoperation of collaborating systems or services at the identified linguistic levels (first and second columns).

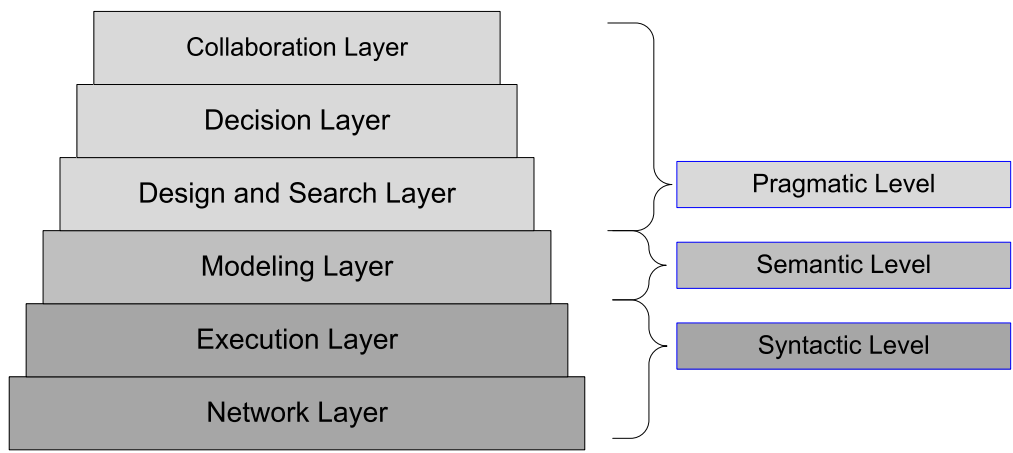

**Figure 3** Associating Linguistic Levels with Layers of Modeling and Simulation

| Linguistic Level | A collaboration of systems or services interoperates at this level if: | Examples |
|---|---|---|
| Pragmatic – how information in messages is used | The receiver reacts to the message in a manner that the sender intends | An order from a commander is obeyed by the troops in the field as the commander intended. A necessary condition is that the information arrives in a timely manner and that its meaning has been preserved (semantic interoperability) |
| Semantic – shared understanding of meaning of messages | The receiver assigns the same meaning as the sender did to the message. | An order from a commander to multi-national participants in a coalition operation is understood in a common manner despite translation into different languages. Similarly geographic data must be translated correctly to UTM grid coordinates for ground forces and to LatLong for air and naval forces. |
| Syntactic – common rules governing composition and transmission of messages | The consumer is able to receive and parse the sender's message | A common network protocol (e.g. IPv4) is employed ensuring that all nodes on the network can send and receive data bit arrays adhering to a prescribed format. |

**Table 2.** Linguistic levels of Interoperability

## 5. Review of M&S foundational framework

The theory of modeling and simulation presented in [16] provides a conceptual framework and an associated computational approach to methodological problems in M&S. The framework provides a set of entities and relations among the entities that, in effect, present a ontology of the M&S domain. The computational approach is based on the mathematical theory of systems and works with object orientation and other computational paradigms. It is intended to provide a sound means to manipulate the framework elements and to derive logical relationships among them that are usefully applied to real world problems in simulation modeling. The framework entities are formulated in terms of the system specifications provided by systems theory, and the framework relations are formulated in terms of the morphisms (preservation relations) among system specifications. Conversely, the abstractions provided by mathematical systems theory require interpretation, as provided by the framework, to be applicable to real world problems.

In its computational realization, the framework is based on the DEVS formalism and implemented in various object oriented environments. Using Unified Modeling Language (UML) we can represent the framework as a set of classes and relations as illustrated in Figures 4 and 5. The Framework for M&S as described in [16] establishes entity classes that are: source system, model, ontology, simulator, and experimental/pragmatic frames. These classes are related by the modeling and the simulation relationships. Each entity is formally characterized as a system at an appropriate level of specification of a generic dynamic system. The source system is the real or virtual environment that we are interested in modeling. It is viewed as a source of observable data, in the form of time-indexed trajectories of variables. The data that has been gathered from observing or otherwise experimenting with a system is called the system behavior database. This data is viewed or acquired through experimental frames of interest to the model development and user. These data must be sufficient in scope to enable reliable comparison as well accepted by both the model developer and the test agency as the basis for comparison. Data sources for this purpose might be measurement taken in prior experiments, mathematical representation of the measured data, or expert knowledge of the system behavior by accepted subject matter experts. An experimental frame is a specification of the conditions under which the system is observed or experimented with. An experimental frame is the operational formulation of the objectives that motivate a M&S project. A frame is realized as a system that interacts with the system of interest to obtain the data of interest under specified conditions. When an experimental frame is realized as a system to interact with the model or system under test the specifications become components of the driving system. Pragmatic frames were recently introduced in [11] to generalize the concept of experimental frame to represent the objectives involved in creating ontologies. System specification morphisms are implemented as relationships among entity classes. For example, the validity of a model with respect to a source system is characterized through a morphism at the behavioral level and implemented as a relationship between pairs of model and source system instances. Various implementations support different subsets of the classes and relations [OMG]. In particular, this article will review the implementation of DEVS within a Service Oriented Architecture (SOA) environment called DEVS/SOA [17,33,34].

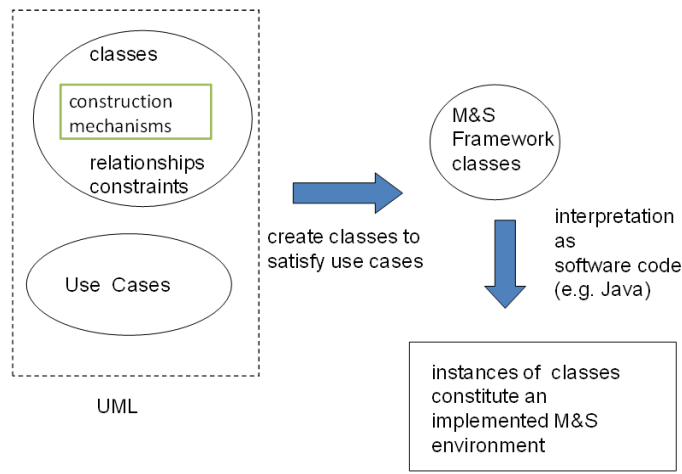

**Figure 4:** M&S Framework formulated within UML

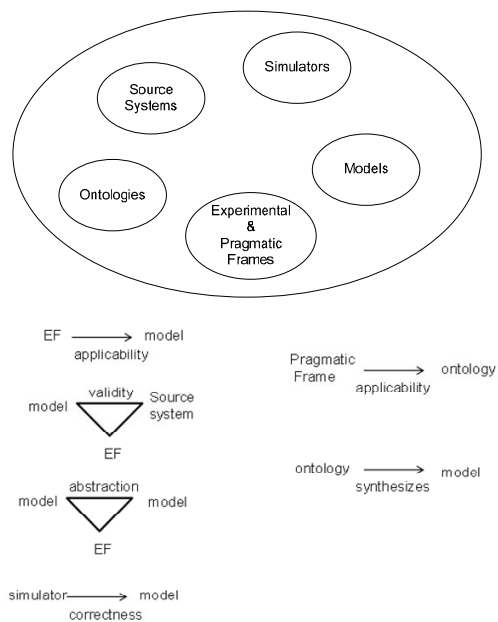

**Figure 5:** M&S Framework Classes and Relations in a UML representation

In a System of systems, systems and/or subsystems often interact with each other because of interoperability and over all integration of the SoS. These interactions are achieved by efficient communication among the systems using either peer-to-peer communication or through central coordinator in a given SoS. Since the systems within SoS are operationally independent, interactions among systems are generally asynchronous in nature. A simple yet robust solution to handle such asynchronous interactions (specifically, receiving messages) is to throw an event at the receiving end to capture the messages from single or multiple systems. Such system interactions can be represented effectively as discrete-event models. In discrete-event modeling, events are generated at random time intervals as opposed to some pre-determined time interval seen commonly in discrete-time systems. More specifically, the state change of a discrete-event system happens only upon arrival (or generation) of an event, not necessarily at equally spaced time intervals. To this end, a discrete-event model is a feasible approach in simulating the SoS framework and its interaction. Several discrete-event simulation engines [35-38] are available that can be used in simulating interaction in a heterogeneous mixture of independent systems. The advantage of DEVS is its effective mathematical representation and its support to distributed simulation using middleware such as DoD's HLA [39].

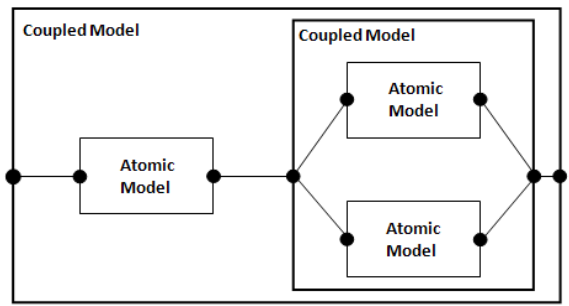

**Figure 6:** DEVS Hierarchical Model representation for systems and sub-systems

## 5.1 DEVS Modeling and Simulation

DEVS [16] is a formalism, which provides a means of specifying the components of a system in a discrete event simulation. In DEVS formalism, one must specify *Basic Models* and how these models are connected together. These basic models are called *Atomic Models* and larger models which are obtained by connecting these atomic blocks in meaningful fashion are called *Coupled Models* (shown Figure 6). Each of these atomic models has *inports* (to receive external events), *outports* (to send events), set of *state variables*, *internal transition*, *external transition*, and *time advance functions*. Mathematically it is represented as 7-tuple system: $M = <X, S, Y, \delta_{int}, \delta_{ext}, \lambda, t_a>$ where $X$ is an input set, $S$ is the set of states, $Y$ is the set of outputs, $\delta_{int}$ is the internal transition function, $\delta_{ext}$ is the external transition function, $\lambda$ is the output function, and $t_a$ is the time advance function. The model's description (implementation) uses (or discards) the message in the event to do the computation and delivers an output message on the outport and makes a state transition. A Java-based implementation of DEVS formalism, DEVSJAVA [40], can be used to implement these atomic or coupled models. In addition, DEVS-HLA [40] will be helpful in distributed simulation for simulating multiple heterogeneous systems in the System of systems framework.

DEVS formalism categorically separates the Model, the Simulator and the Experimental frame (Figure 7). However, one of the major problems in this kind of mutually exclusively system is that the formalism implementation is itself limited by the underlying programming language. In other words, the model and the simulator exist in the same programming language. Consequently, legacy models as well as models that are available in one implementation are hard to translate from one language to another even though both the implementations are object oriented. Other constraints like libraries inherent in C++ and Java are another source of bottleneck that prevents such interoperability.

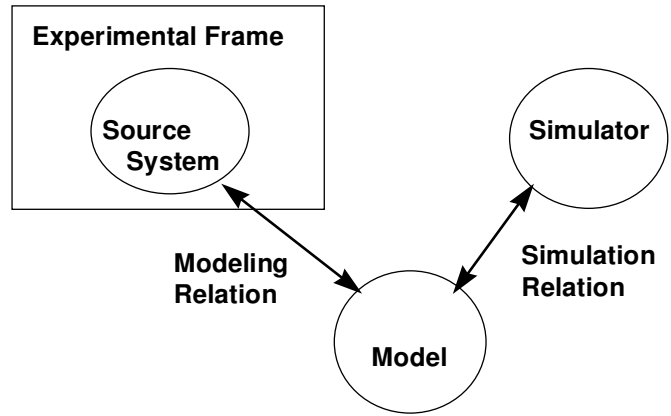

**Figure 7:** Framework Entities and Relationships

**Brief Overview of Capabilities Provided by DEVS**

The prime motivation comes from an editorial by Carstairs [41] that demands a M&S framework at higher levels of system specifications where System of systems interact together using net-centric platform. At this level, model interoperability is one of the major concerns. The motivation for this work stems from this need of model interoperability between the disparate simulator implementations and provides a means to make the simulator transparent to model execution. DEVS, which is known to be component-based system, based on formal systems theoretical framework is the preferred means. Table 3 outlines how it could provide solutions to the challenges in net-centric design and evaluation. The net-centric DEVS framework requires enhancement to the basic DEVS capabilities, which are provided in later sections.

| Desired M&S Capability for Test and Evaluation (T&E) | Solutions Provided by DEVS Technology |
|---|---|
| Support of DoDAF need for executable architectures using M&S such as mission based testing for GIG/SOA | DEVS Unified Process [33,42] provides methodology and SOA infrastructure for integrated development and testing, extending DoDAF views [21]. |
| Interoperability and cross-platform M&S using GIG/SOA | Simulation architecture is layered to accomplish the technology migration or run different technological scenarios [43]. Provide net-centric composition and integration of DEVS 'validated' models using Simulation Web Services [26] |
| Automated test generation and deployment in distributed simulation | Separate a model from the act of simulation itself, which can be executed on single or multiple distributed platforms [16]. With its bifurcated test and development process, automated test generation is integral to this methodology [45]. |
| Test artifact continuity and traceability through phases of system development | Provide rapid means of deployment using model-continuity principles and concepts like "simulation becomes the reality" [29]. |
| Real time observation and control of test environment | Provide dynamic variable-structure component modeling to enable control and reconfiguration of simulation on the fly [47-49]. Provide dynamic simulation tuning, interoperability testing and benchmarking. |

**Table 3:** Solutions provided by DEVS technology to support of M&S for T&E

Furthermore, this work describes distributed simulation using the web service technology. After the development of World Wide Web, many efforts in the distributed simulation field have been made for modeling, executing simulation and creating model libraries that can be assembled and executed over WWW. By means of XML and web services technology these efforts have entered upon a new phase. The proposed DEVS Modeling Language (DEVSML) [26] is built on eXtensible Markup Language (XML) [50] as the preferred means to provide such transparent simulator implementation. A prototype simulation framework called DEVS/SOA has been implemented using web services technology. It is currently in use by various research groups across the world towards a global net-centric simulation platform [51]. The central point resides in executing the simulator as a web service. The development of this kind of frameworks will help to solve large-scale problems and guarantees interoperability among different networked systems and specifically DEVS-validated models. This paper focuses on the overall approach, and the symmetrical SOA-Based architecture that allows for DEVS execution as a Simulation SOA.

## 6. DEVS Standard

The conceptual interoperability model described above provides a general guideline for supporting system interoperability. Following the layered approach of this conceptual model, next we review the work of DEVS standardization that aims to support M&S interoperability based on the DEVS M&S framework. This work of standardization correspond to the two levels shown in Figure 3: the semantic level that deals with standardization of model interface; and the syntactic level that deals with standardization of simulation protocol.

The DEVS formalism [16], based on Mathematical Systems theory, provides a computational framework and tool set to support Systems concepts in application to SoS. We first provide a brief review. More detail is available in [16].

DEVS makes a sharp distinction between the model and the device that simulates it. Both model and simulator are defined as mathematical systems as defined by Wymore and others (see [16] for details), and the relation between them is standardized by the concept of "abstract" simulator. Information flow in the DEVS formalism, as implemented on an object-oriented substrate, is mediated by the concept of DEVS message, a container for port-value pairs. In a message sent from component A to component B, a port-value pair is a pair in which the port is an output port of A, and the value is an instance of the base class of a DEVS implementation, or any of its sub-classes. A coupling is a four-tuple of the form (*sending component A, output port of A, receiving component B, input port of B*). This sets up a path where by a value placed on an output port of A by A's output function is transmitted to the input port of B, to be consumed by the latter. In systems or simulations implemented in DEVS environments the concepts of ports, messages, and coupling are explicit in the code. However, for systems/simulations that were implemented without systems theory guidance, in legacy or non-DEVS environments, these concepts are abstract and need to be identified concretely with the constructs offered by the underlying environment. For SoS engineering, where legacy components are the norm, it is worth starting with the clear concepts and methodology offered by systems theory and DEVS, getting a grip on the interoperability problems, and then translating backwards to the non-DEVS concepts as necessary.

Within a working group of the Simulation Interoperability Standards Organization, a standard has been under development to support interoperability of DEVS models implemented in different platforms as well as with legacy simulations. Figure 8 illustrates an architectural approach proposed to accommodate the various combinations and permutations of possible application, both currently known, as well as those that will emerge in the future. The basic idea is to define two sets of interfaces; the DEVS model Interface and the DEVS Simulator Interface, as well as a DEVS Simulation Protocol that operates between the two. The interfaces protocols are based on those in GenDEVS, an implementation at the heart of the DEVJAVA M&S environment [www.acims.arizona.edu]. DEVS/C++ and DEVSJAVA are platform specific implementations while DEVSML[26] and XFD-DEVS [27] are platform independent implementations in XML which can transform to any platform specific implementations.

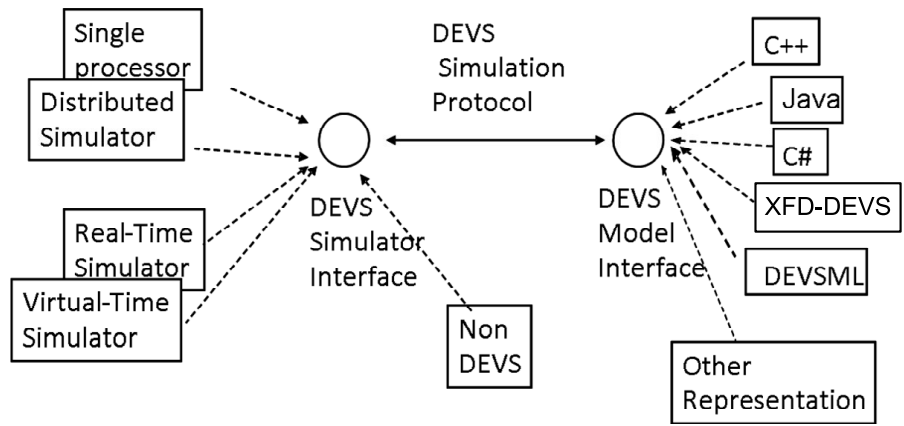

**Figure 8:** Conceptual Architecture of Standard

As a direct consequence of the model-simulator separation there can be multiple ways in which the same model can be simulated – all adhering to the abstract simulator specification. Corresponding to different simulation modes, the standard has virtual-time and real-time simulators. In virtual-time simulation, the simulator interprets time as logical time so the simulation can skip from one event time to the next without traversing the intervening time interval. However, in real-time simulation, time is interpreted as wall clock readings, so the real-time simulator will wait for the interval to its next scheduled event to expire before handling the event. In addition to the model type/simulation mode combinations, the standard allows for the use of different forms of distribution of model components, e.g., single processor vs. multi-processor, and within the latter, conservative vs. optimistic time advance for virtual-time as well as centralized vs. non-centralized time control in real-time execution. The standard is also agnostic with respect to different implementation platforms, such as Windows vs. Unix, different programming languages, such as Java vs. C++, and different networking and middleware frameworks such as .Net vs. Apache. From the above introduction, we can see that the standard will have multiple simulation scenarios. For example, considering the combinations of simulation mode and distribution mode, we have: simulating a model in virtual-time and simulating a model in real-time both in distributed and non-distributed fashion.

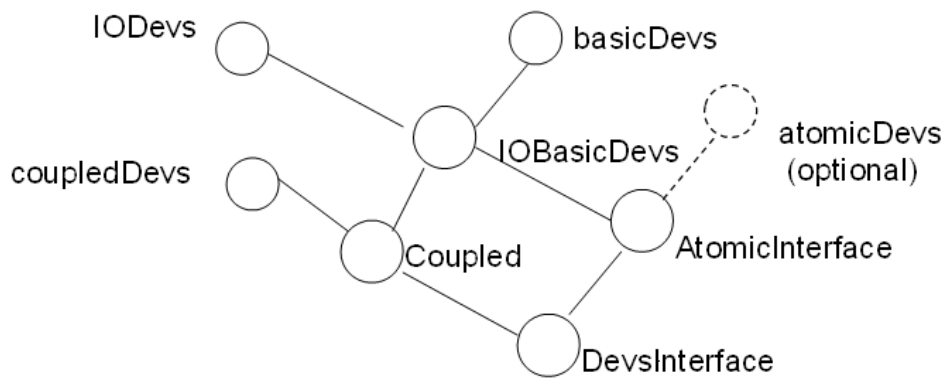

**Figure 9** DEVS Model Interfaces

Among the interfaces (Figure 9), *IODevs* defines interface for the functions that handle message exchange based on input and output ports. Any model, whether DEVS or non-DEVS, can implement these functions so it can interoperate with other implementers of this interface, in the sense of receiving input and sending output. The *basicDevs* Interface defines the basic functions a DEVS model needs to implement such as *deltext()*, *deltint()*, *out()*, *ta()* and so on. The *basicDevs* interface is the interface that is exposed to the atomic simulators. An additional interface, *atomicDevs*, provides a convenient set of primitives for defining the basic functions in an atomic model. However, since the basic functions can be defined without using such primitives, the *atomicDevs* interface is optional. The *IOBasicDevs* interface extends the *IODevs* interface and *basicDevs* interface. It provides a common basis for implementing atomic models and coupled models. Combining *IOBasicDevs* with *atomicDevs*, we get *AtomicInterface* which defines the function signatures an atomic model need to implement. Of course, if *atomicDevs* is omitted, then *AtomicInterface* reduces to *IOBasicDevs*. Similarly, *CoupledDevs* interface defines the function signatures that are used in DEVS coupled models. It also has methods that support adding components and couplings to the model; methods for retrieving a component by name and for accessing all components; and to access the internal coupling specifications (intended only by simulators). Combining *IOBasicDevs* with *CoupledDevs*, we get the Coupled interface which defines the functions coupled models need to implement.

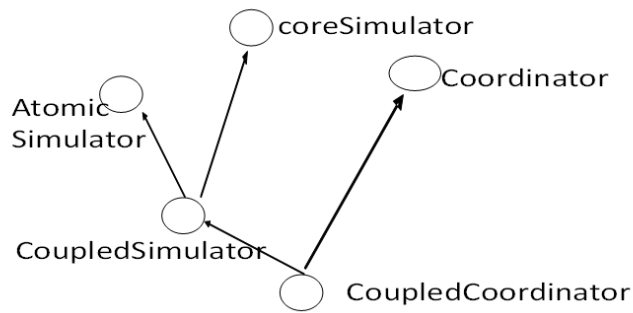

**Figure 10** DEVS Simulator Interfaces

The basic simulator interface is the *CoreSimulator* that provides a common interface for DEVS and non-DEVS simulation (Figure 10). Further, the *CoreSimulator* interface is *the* basic interface from which simulation services could be designed for a truly net-centric interoperable simulation framework [23]. Under the *CoreSimulator* interface, two classes of simulators have been defined *CoupledSimulator* and *CoupledCoordinator* interfaces where the latter also inherits from *Coordintor*. These apply to both virtual (logical); and real-time simulation. (Real time simulators interpret time as real wall clock time and have their own thread and system clock. Virtual or logical time simulators can advance from one event time to the next). The *CoreSimulator* interface includes methods that are invoked by the DEVS simulation protocol:

>  *interface coreSimulatorInterface{*
>  
>  *void setSimulators (Collection<CoreSimulatorInterface>);*
>  
>  *void initialize();*
>  
>  *Double nextTN();*
>  
>  *void computeInputOutput(Double t);*
>  
>  *void applyDeltFunc(Double t);*
>  
>  *void putContentOnSimulator (CoreSimulatorInterface sim, ContentInterface c);*
>  
>  *void sendMessages();*

## 6.1 DEVS Simulation Protocol

DEVS treats a model and its simulator as two distinct elements. The simulation protocol describes how a DEVS model should be simulated whether in standalone fashion or in a coupled model. Such a protocol is implemented by a processor which can be a simulator or a coordinator.

As illustrated in Figure 11, the DEVS protocol is executed as following:

1. It starts with the coordinator telling each of the simulators in the collection the others' addresses and then to perform initialization function.
2. A cycle is then entered in which the coordinator requests that each simulator provide its time of next event and takes the minimum of the returned values to obtain the global time of next event
3. Each of the simulators applies its *computeInputOutput*() method to produce an output that consists of a collection of contents (port/value) pairs – for DEVS simulators this is a composite message computed according to the DEVS formalism based on its model's current state.
4. Then each simulator partitions its output into messages intended for recipient simulators and sends these messages to these recipient simulators – for DEVS simulators these recipients are determined from the output ports in the message and the coupling information that will have previously been received from the coordinator.
5. Finally, each simulator executes its *ApplyDeltFunc* method which computes the combined effect of the received messages and internal scheduling on its state, a side effect of which is produce of time of next event, *tN* – for DEVS simulators this state change is computed according to the DEVS formalism and the *tN* is updated using its model's time advance.
6. The coordinator obtains the next global time of next event and the cycle repeats

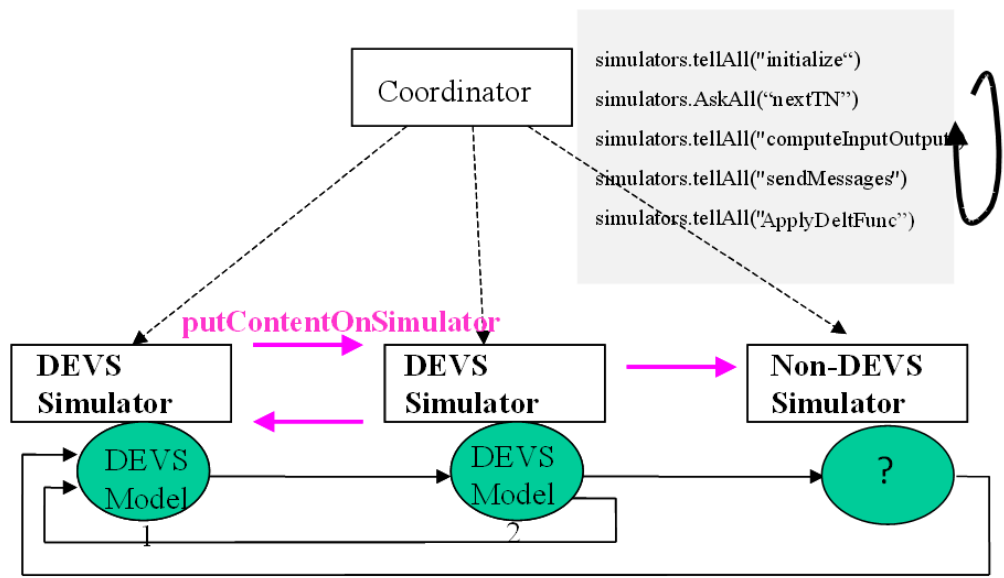

**Figure 11.** Federation of DEVS with Non-DEVS Simulators

It should be noted that the above is one form of many possible protocols that can provide various forms of conservative and optimistic simulation, each of which must be proved to be correct as a realization of the DEVS closure under coupling property [16]. One such implementation exists as Microsim/Java [44] wherein the DEVS simulation protocol adheres to the *CoreSimulator* interface but has different implementation when compared to GenDEVS.

Implicit in the above description are the following constraints involving methods in the *CoreSimulatorInterface*:

- The *sendMessages*() method "must" employ the *putContentOnSimulator*() method as follows: for any simulator to which it wishes to send a content, it must call the recipient's *putContentOnSimulator*() method with the recipient and the content as arguments.

- Further, in applying its *computeInputOutput*() method, a simulator "must" be able to interpret the contents (satisfying the ContentInterface) it has received from the other simulators.

Notice that we cannot enforce the "must" requirements just given, and cannot prove that the simulation executes a desired behavior, unless we are given further information about its behavior. One way to do this is where the simulators are truly DEVS simulators in that they satisfy the interfaces and constraints given below. Failing this additional rigor, the interoperation involving DEVS and non-DEVS is purely at the technical level similar to that of a federation of simulators in HLA. This contrasts with the situation in which the federation is in fact derived from a DEVS coupled model for which correct simulation of the coupled model is guaranteed according to the DEVS formalism.

## 7. DEVS/SOA

An implementation of the standard within the Service Oriented Architecture (SOA) has been completed that provides DEVS modeling and simulation services over the World Wide Web [17, 23]. As shown in the Figure 12, at the top of the layered architecture is the application layer that contains models in DEVSJAVA or DEVSML, a way of representing DEVS models in the eXtended Markup Language (XML). This DEVSML is built on JAVAML [18], which is XML implementation of JAVA. The current development effort of DEVSML takes its power from the underlying JAVAML that is needed to specify the 'behavior' logic of atomic and coupled models. The DEVSML models are cross-transformable to Java. The second layer is the DEVSML layer itself that provides seamless integration, composition and dynamic scenario construction resulting in portable models in DEVSML that are complete in every respect. These DEVSML models can be ported to any remote location using the SOA infrastructure and can be executed at any remote location in a distributed or non-distributed manner. Another major advantage of such capability is total simulator 'transparency'. The simulation engine is totally transparent to model execution over the SOA infrastructure. The DEVSML model description files in XML contains meta-data information about its compliance with various simulation 'builds' or versions to provide true interoperability between various simulator engine implementations. This has been achieved for at least two independent simulation engines as they have an underlying DEVS protocol to adhere to. This has been made possible with the implementation of a single atomic schema [24] and a single coupled schema [25] that validates the DEVSML descriptions generated from these two implementations. Such run-time interoperability provides great advantage when models from different repositories are used to compose large coupled models using the DEVSML integration capabilities. Detailed design can be seen in [17,23].

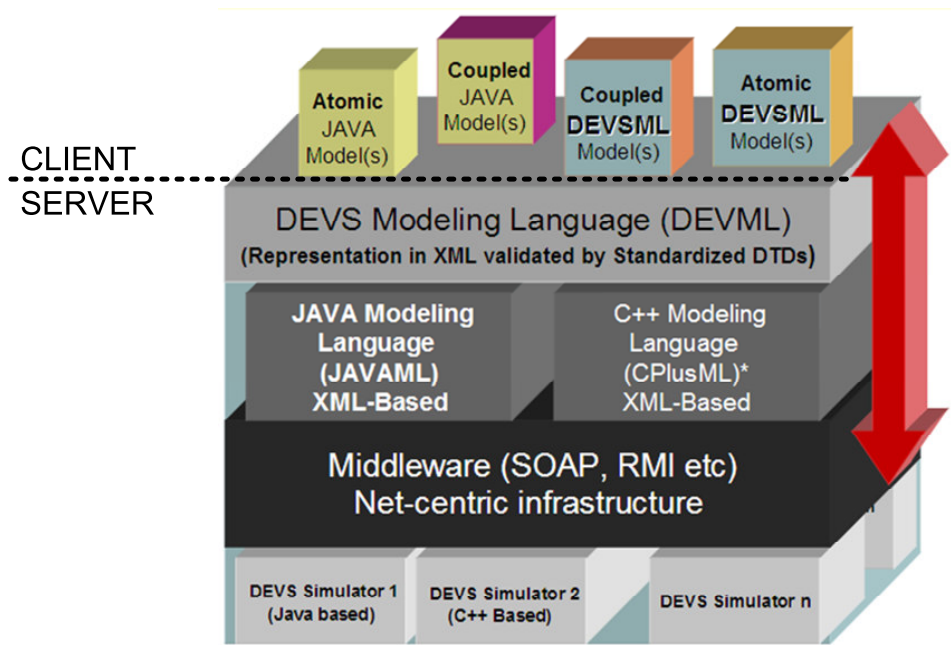

**Figure 12:** Layered Architecture of DEVSML towards transparent simulators in Net-centric domain

The complete setup requires one or more servers that are capable of running DEVS Simulation Service, as shown in Figure 12 by the dotted line. The capability to run the simulation service is provided by the server side design of DEVS Simulation protocol supported by the DEVSJAVA and Microsim/Java. Of course, many issues of policy management and security considerations must be taken care of in the generation of DEVS models from WSDLs specifications [22]. Furthermore, the multi-platform simulation capability provided by DEVS/SOA framework consists of realizing distributed simulation among different DEVS platforms or simulator engines such as DEVSJAVA, Microsim/Java, DEVS-C++, etc. and executing the native simulation service. This kind of interoperability where multi-platform simulations can be executed with our DEVSML integration facilities has been made possible with the hierarchical design of simulator interfaces as described in Section 6.

Web-based simulation requires the convergence of simulation methodology and WWW technology (mainly Web Service technology). The fundamental concept of web services is to integrate software application as services. Web services allow the applications to communicate with other applications using open standards. We are offering DEVS-based simulators as a web service, and they must have these standard technologies: communication protocol (Simple Object Access Protocol, SOAP), service description (Web Service Description Language, WSDL), and service discovery (Universal Description Discovery and Integration, UDDI). Figure13 shows the framework of the proposed distributed simulation using SOA.

The Simulation Service framework is two layered framework as depicted in Figure 13. The top-layer is the user coordination layer (*MainService*) that oversees the lower layer (SimulationService). The lower layer is the true simulation service layer that executes the DEVS simulation protocol as a Service. The lower layer is transparent to the modeler and only the top-level is provided to the user.

The top-level (*MainService* layer) has four main services:

- Upload DEVS model
- Compile DEVS model
- Simulate DEVS model (centralized)
- Simulate DEVS model (distributed)

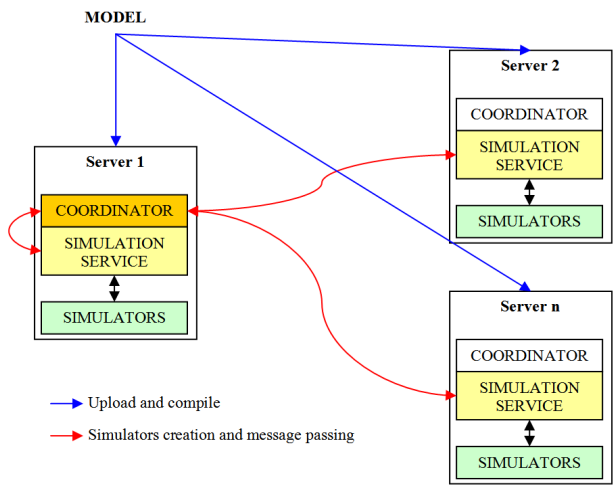

**Figure 13:** DEVS/SOA distributed architecture

The second lower (*SimulationService*) layer provides the DEVS Simulation protocol and is designed as per the DEVS Standard described earlier:

- Initialize simulator i
- Run transition in simulator i

- Run lambda function in simulator i
- Inject message to simulator i
- Get time of next event from simulator i
- Get time advance from simulator i
- Get console log from all the simulators
- Finalize simulation service

The explicit transition functions, namely, the internal transition function, the external transition function, and the confluent transition function, are abstracted to a single transition function that is made available as a Service. The transition function that needs to be executed depends on the simulator implementation and is decided at the run-time. For example, if the simulator implements the Parallel DEVS (P-DEVS) formalism, it will choose among internal transition, external transition or confluent transition[2].

The client is provided a list of servers hosting DEVS Service. He selects some servers to distribute the simulation of his model. Then, the model is uploaded and compiled in all the servers. The main server selected creates a coordinator that creates simulators in the server where the coordinator resides and/or over the other servers selected.

Summarizing from a user's perspective, the simulation process is done through three steps (Figure 14):

1. Write a DEVS model (currently DEVSJAVA is only supported).
2. Provide a list of DEVS servers (through UDDI, for example). Since we are testing the application, these services have not been published using UDDI by now. Select N number of servers from the list available.
3. Run the simulation (upload, compile and simulate) and wait for the results.

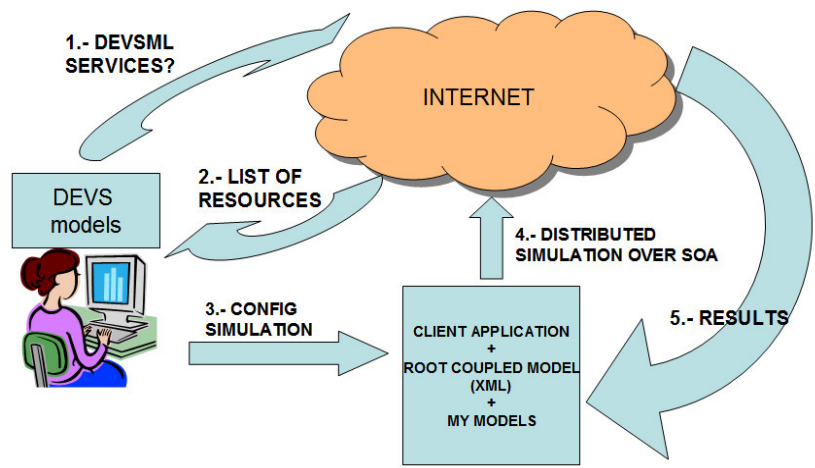

**Figure 14:** Execution of DEVS SOA-Based M&S

## 7.1 Simulation service composition:
**This is the bottom layer of the two-layer** architecture and its functionalities are used by the *MainSevice* layer. Its operations are transparent to the user. Once the user demands a simulation via the *MainService* class, the coordinator (at the coordinator server or main server) requires as many simulation services as IP addresses provided by the user. After that, the DEVS model is partitioned and the coordinator sends every part to its corresponding service. When the simulation starts, each simulation service creates a DEVS simulator for its models and executes the corresponding output and transition functions (see Figure 14).

---
[2] The difference between P-DEVS and classic DEVS is the handling of confluent function. The DEVS/SOA framework could have been built using other simulation formalisms. In fact, our simulation services could store any kind of simulator -as long as the service updates the simulation cycle according to the simulator engine selected. The service is independent in the sense of transition functions.

It is possible for one simulation service to store more than one simulator for different components of the same DEVS model, or to store more than one simulator for different components of different DEVS models. This issue is solved as follows. After the main coordinator obtains a simulation service at a certain IP address, a new simulator is created there, identified by the component name plus the IP address of the user's machine and containing the DEVS component itself. For example, if the coordinator must send a DEVS component named *Processor* to a server located at 192.168.1.5 and coming from a user located at 192.168.1.2, then a simulation service is required from 192.168.1.5 and a new simulator is created there, identified by *Processor@192.168.1.2* and containing the model named *Processor*.

Another issue is how to store the simulators created, because web services do not have memory. To this end, we are using the server's memory by means of static variables or attributes. Hence, the simulation services include a static table, which associates *simulator names* with *simulator instances*. There is other information stored by the Simulation services in the server memory, such as the IP address where the services reside and a reporter, which logs all the information while the simulation is running.

The services provided by the Simulation service are enumerated below in detail:

- *newSimulator:* This service receives a DEVS component and a identifier. It creates a new DEVS simulator identified by the name described above and containing the DEVS component received.
- *initialize:* This service receives the name of the simulator required and the current time. It takes the corresponding simulator from its table (using the name received) and initializes it.
- *receiveInput:* This service receives four arguments: (1) the name of the simulator required, (2) the name of the port where the message is coming from, (3) the message and (4) the name of the port where the message is going to. The simulation service takes the simulator from its table and executes the same function called *receiveInput*, which stores the message received at the input of the model.
- *lambda:* It receives the name of the simulator required and the current time. This service takes the simulator required and executes the output function (also called *lambda)* of the DEVS model
- *deltfnc:* This service receives the name of the simulator required and the current simulation time. The service takes the simulator and executes an internal or external or confluent transition function. The abstracted *deltfn* is provides in Figure 15. This allows both the classical DEVS and P-DEVS models work seamlessly with DEVS/SOA simulation framework.
- *getOutput:* This service takes the required Simulator and returns the output stored in its DEVS model.
- *getTN:* It receives the name of the simulator for which the time of the next event is returned.
- *exit:* It receives the name of the simulator to be removed from the table.
- *getConsole:* This service receives the IP address of the user's machine, and return the content of the log file related to this address.
- *getIp:* It returns the IP address of the simulation service.

Having described the services available in the DEVS/SOA architecture, following is the design of DEVS/SOA coordinator and simulator that utilize these DEVS services. This simulator is called as DEVSV/SOA simulator and it acts as an adapter for any DEVS simulation engine that executes the DEVS simulation protocol. Currently, we have implemented the DEVS/SOA simulator in two DEVS implementations viz. DEVSJAVA and Microsim/Java. More details on a complete example can be seen in [23] that use these two independent implementations of abstract simulator interface. Efforts are underway for a DEVS.net implementation using C# language.

```
function deltfcn(double t) {
        Message x = input;
        if(x==null) {
                System.out.println(
                "ERROR RECEIVED NULL INPUT " + model.toString());
                return;
        }

        //if you receive an empty message and not imminent
        if (x.isEmpty() && t!=tN) {
                return;
        }

        //if incoming message is not empty and imminent
        //update the elapsed time, sigma
        //execute the deltcon transition function
        else if((!x.isEmpty()) && t==tN) {
                double e = t – tL;
                model.deltcon(e,x);
        }

        //if just imminent and no message
        //execute deltint transition function
        else if(t==tN) {
                model.deltint();
        }

        //if not imminent and just a message incoming
        //execute deltext transition function
        else if(!x.isEmpty()) {
                double e = t – tL;
                model.deltext(e,x);
        }

        //update tL (time of last event) and tN (time of next event)
        //update sigma (time advance)
        //reset incoming message collection
        tL = t;
        tN = tL + model.ta();
        input = new Message();
}
```

**Figure 15:** Abstract *deltfun* in Simulation service

### **DEVSV/SOA Coordinator**

Equivalent to the Simulation service storing the simulators in a static way, the coordinator also stores the simulators of the DEVS model in a static hash table, using the same nomenclature as was stated above (DEVS component name plus client IP address identifying the simulator). Therefore, such table contains pairs {simulator name, simulator service}, associating each simulator created with the simulation service where it resides. The task of the coordinator is to execute a typical DEVS loop over the distributed simulators. Figure 16 shows the algorithm executed by the *simulate* function. In such Table, *iterations* is the number of cycles of the simulation, *t* is the current time, *tL* is the last time event, *tN* is the next time event, *simulationServices* is the table of simulation services created by the coordinator and where the simulators are located. Then, for a number of cycles, the output function is called through each of the simulation services. It should be noted that the first argument of *lambda* function is a key, which is the simulator identifier, since different simulators could be located at the same simulation service, this key must be provided. After the output function is executed, the outputs of the components are ready to be propagated. To this end, the *propagateOutput* function is called, which propagates the messages generated from the outports to its corresponding inports. Next, the transition function is applied and finally the time is updated.

From the instant in which the coordinator is created, it stores at any moment the DEVS model (currently DEVSJAVA), the last timed event, the next time event and the IP address of the user's machine.

```
function simulate(long iterations)
  t = tN;
  for (i=0; i<iterations; i++)
    for each ({key,simService} in simulationServices)
      simService.lambda(key, t);
      propagateOutput();
    for each ({key,simService} in simulationServices)
      simService.deltfcn(key, t);
    tL = t;
    tN = min(simulationServices.getTN());
    t = tN;
```

**Figure 16:** DEVS simulation

It should be noted that the Coordinator is not a service. It is a class, which is used by the *MainService* service. Again, it must be stressed that it derives from the DEVS Standard simulator interface, which allows a DEVS coordinator to control and coordinate a DEVS simulator. The DEVS Standard interface has been extended for obvious reasons and extended functionalities. The functions implemented in the Coordinator are enumerated below:

- *getTopComponentNames:* This function receives the name of the DEVS root-coupled model and returns a list containing the top-component names of the DEVS model.
- *Constructor:* The constructor receives the client IP address, the name of the DEVS model, and the list of IP addresses where the model is going to be simulated. Hence, it creates as many simulators as top-level components, created by the simulation services located at the IP addresses given in the list.
- *initialize:* This function receives the initial time of simulation. It initializes the simulators.
- *propagateOutput:* As it was stated above, this function takes the output from the simulators and sends them to its corresponding inputs.
- *lamda:* It receives the current time, and executes the output function in each of the simulators stored.
- *deltfcn:* This function receives the current time and executes the internal or external transition functions in the simulators stored.
- *ta:* It is the time advance function and receives the current time. It takes the minimum next time event from the simulators stored.
- *exit:* This function calls the exit function of all the simulation services stored and clean the table of simulators.
- *simulate:* This function receives the number of cycles of the simulation, and executes the simulation as was described before.

## 7.2 Client Application

This Section provides the client application to execute DEVS model over an SOA framework using Simulation as a Service. From many-sided modes of DEVS model generation [33,34], the next step is the simulation of these models. The DEVSV/SOA client takes the DEVS models package and through the dedicated servers hosting simulation services, it performs the following operations:

1. Upload the models to specific IP locations
2. Run-time compile at respective sites
3. Simulate the coupled-model
4. Receive the simulation output at client's end

The DEVSV/SOA client as shown in Figure 17 operates in the following sequential manner:

1. The user selects the DEVS package folder at his machine
2. The top-level coupled model is selected as shown in Figure 17.
3. Various available servers are selected (Figure 17). Any number of available servers can be selected (one at least).

4. Clicking the button labelled "Assign Servers to Model Components" the user selects where is going to simulate each of the coupled models, including the top-level one, i.e., the main server where the coordinator will be created (Figure 18)
5. The user then uploads the model by clicking the Upload button. The models are partitioned and distributed among the servers chosen in the previous point
6. The user then compiles the models at the server's end by clicking the Compile button

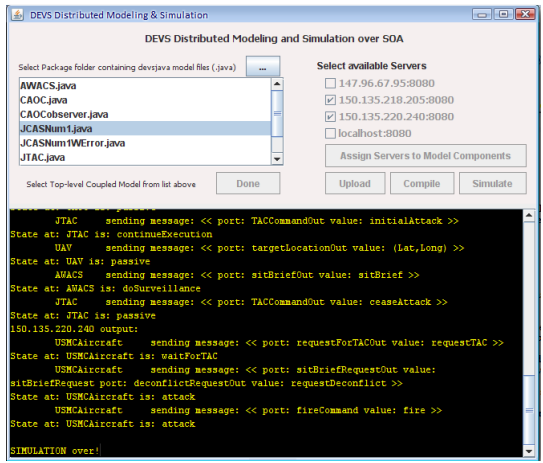

**Figure 17:** GUI snapshot of DEVSV/SOA client hosting distributed simulation

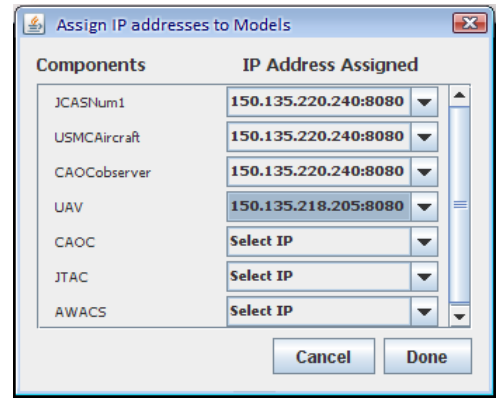

**Figure 18:** Server Assignment to Models

## 8. Cross-Platform Execution over DEVS/SOA

In terms of net-ready capability testing, what is required is the communication of live web services with those of test-models designed specifically for them. The approach we are working on has the following steps:

1. Specify the scenario
2. Develop the DEVS model
3. Develop the test-model from DEVS models
4. Run the model and test-model over SOA
5. Execute as a real-time simulation
6. Replace the model with actual web-service as intended in scenario.
7. Execute the test-models with real-world web services
8. Compare the results of steps 5 and 7.

Of course, many issues of policy management and security considerations must be taken care of when test-models are communicating with live Web-Services. However, considering the fact that for any defense related mission-thread reliability testing the test-models would have the necessary security provisions, the 8-step process listed above can be executed. This work would also involve generation of DEVS models from Web Service Description Language or WSDLs specifications. A small portion of Business Process Modeling Notation (BPMN) to DEVS transformation is described in [33].

One other section that requires some description is the multi-platform simulation capability as provided by DEVSV/SOA framework. It consists of realizing distributed simulation among different DEVS platforms or simulator engines such as DEVSJAVA, DEVS-C++, etc. In order to accomplish that, the simulation services will be developed that are focused on specific platforms, however, managed by a coordinator. In this manner, the whole model will be naturally partitioned according to their respective implementation platform and executing the native simulation service. This kind of interoperability where multi-platform simulations can be executed with our DEVSML integration facilities. DEVSML will be used to describe the whole hybrid model. At this level, the problem consists of message passing, which has been solved in this work by means of an adapter pattern in the design of the "message" class [23]. Figure 19 shows a first approximation. The platform specific simulator generates messages or events, but the simulation services will transform these platform-specific-messages (PSMsg) to our current platform-independent-message (PIMsg) architecture developed in DEVS/SOA. Hence, we see that the described DEVS/SOA framework can be extended towards net-ready capability testing. The DEVS/SOA framework also needs to be extended towards multi-platform simulation capabilities that allow test-models be written in any DEVS implementation (e.g. Java and C++) to interact with other as services.

However, a major drawback of our current architecture is that the user must send the whole DEVS model implemented under all the platforms to use, which is not a good solution. Next, we propose a modification on the Coordinator creation process that in some manner, allows to the user to store each part of the model written in its corresponding platform.

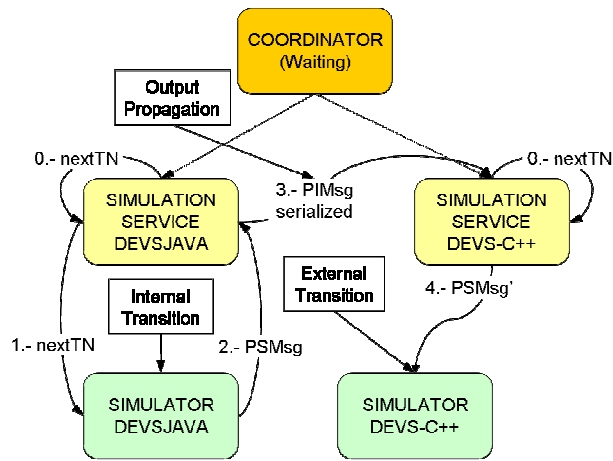

**Figure 19:** Cross-platform execution.

## 8.1 Multi-platform DEVS/SOA architecture

Figure 20 depicts an example of a multi-platform DEVS model. Each atomic or coupled component may be implemented using different simulation engines, called *platforms*. In Figure 20, SUBMODEL A is implemented using DEVSJAVA [28], SUBMODEL B by means of aDEVS (C++) [52], and SUBMODEL C using xDEVS (Java) [38].

Let us suppose that the whole model is implemented using DEVSJAVA. In our current DEVS/SOA architecture, the application sends the whole model (root-coupled model included) to the servers by means of the *upload* service, where all the files get compiled and finally, it executes the model sending serialized messages among simulation services. This situation is not valid for the multi-platform model depicted in

Figure 20 as the scenario cannot be compiled as a whole.

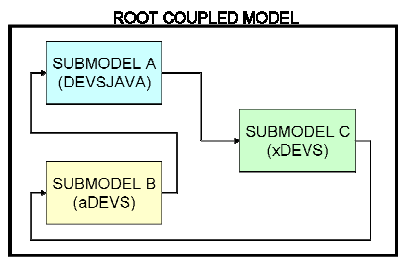

**Figure 20:** Multi-platform DEVS model

In our proposed approach, we define the root coordinator by means of a *Platform Independent Model (PIM)*, for example, DEVSML. We may use the structure description of DEVSML to compose the root coupled model, and send it to the main server, which will distribute the sub-models among its corresponding servers. Figure 21 shows how a multi-platform DEVS model may be executed using our proposed architecture. We define the root-coupled model using DEVSML (top of the Figure 21). The coupled model is treated as an atomic model due to the inherent architecture of DEVS/SOA *digraph2Atomic* adapter [23]. Consequently, it is immaterial if the sub-model is atomic or coupled.

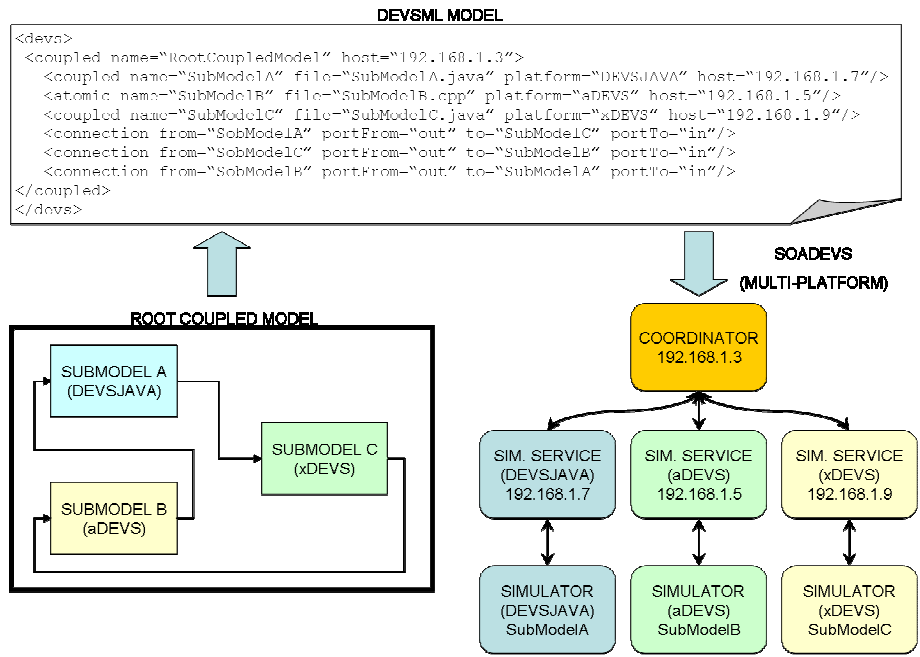

**Figure 21:** Multi-platform DEVSV/SOA proposed architecture

The DEVSML document in the Figure 21 states that the main server is located at 192.168.1.3. This server receives the DEVSML document and all the source code, distributes sub-models to respective servers and creates the coordinator. For example, the main server sends *SubModelA.java* to the server located at 192.168.1.7, where the DEVS/SOA java implemented server compiles it. The same happens with the corresponding *SubModelB.cpp* and *SubModelC.java*. After compiling all sub-models, the main server creates one simulation service for each sub-model. Figure 21 (right side) shows how coordinator, simulation services, and simulators are created. The main server creates a DEVSJAVA-based simulation service located at 192.168.1.7, which also creates a DEVSJAVA-based simulator to store *SubModelA*. The same occurs with sub-models B and C, but at IP addresses 192.168.1.5 and 192.168.1.9 respectively.

The rest of the behavior of the application is the same that in our current architecture. Messages are passed by means of an adapter pattern, which as Figure 19 depicts, may be translated into different platforms.

# 9. How Interoperability is supported
The proposed DEVS standard and its DEVS/SOA implementation support several modes of interoperability. These are outlined in the following paragraphs.

## 9.1 DEVS-to-DEVS Interoperability
DEVS-to-DEVS Interoperability is the basic form of interoperability enabled by the DEVS standard as discussed above. Adoption of the DEVS standard facilitates new development to achieve interoperability at the syntactic, semantic and pragmatic levels mentioned above. More detail on these concepts in application to testing of SOA systems can be found in [5, 20, 21, 22].

## 9.2 DEVS-to-Non-DEVS Interoperability
### 9.2.1 Direct

As mentioned before, legacy simulations that can be refactored to implement the *CoreSimulator* interface can be interoperate at the syntactic level with DEVS and other non-DEVS peers. In its strongest form, such simulation methodology guarantees well-defined time preservation and simulation correctness as a sound basis to aim for interoperability at the higher levels.

### 9.2.2 Via Client Gateways.

For a variety of reasons, although DEVS compliance is desirable, it can be expected that legacy systems will continue to prevail and new non-compliant systems developed. The adoption of the SOA standard however, will facilitate the interoperation of DEVS and non-DEVS components that are compliant with the SOA standard. This form is realized in an Agent-implemented Test Instrumentation Infrastructure that deploys DEVS models to act as agents that are attached to clients of services [5,22]. Such attachment can be performed in automated fashion using tools such as Axis Toolkit to create the client stub given a service's Web Service Description Language (WSDL) [22,53]. As in Figure 22, these agents can observe the web service requests originating from the client and server responses (or failure thereof) to accumulate a variety of performance measurements. The agents can also serve as virtual users to interact with other users to direct the course of test scenarios and collect performance metrics to support scalability studies. Further, while collecting data, DEVS agents can communicate with each other to coordinate and share information using the DEVS-to-DEVS configuration just discussed. Case studies are available in reference [22].

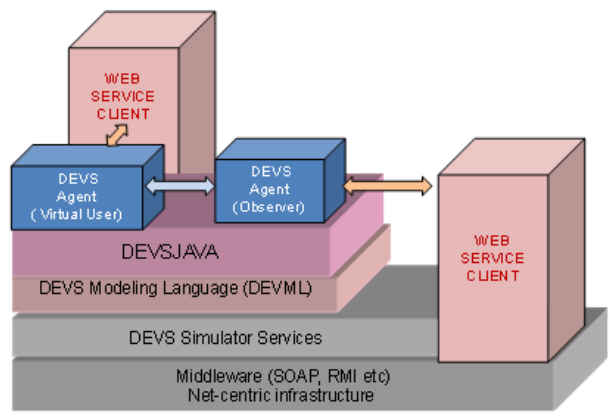

**Figure 22.** DEVS/SOA interoperability

## 10. Multi-layered agent based TEST INSTRUMENTATION system using GIG/SOA

A DEVS distributed federation is a DEVS coupled model whose components reside on different network nodes and whose coupling is implemented through middleware connectivity characteristic of the environment, e.g., SOAP for GIG/SOA. The federation models are executed by DEVS simulator nodes that provide the time and data exchange coordination as specified in the DEVS abstract simulator protocol.

As discussed earlier, in the general concept of experimental frame (EF), the generator sends inputs to the SoS under test (SUT), the transducer collects SUT outputs and develops statistical summaries, and the acceptor monitors SUT observables making decisions about continuation or termination of the experiment [18]. Since the SoS is composed of system components, the EF is distributed among SoS components, as illustrated in Figure 23**.** Each component may be coupled to an EF consisting of some subset of generator, acceptor, and transducer components. As mentioned, in addition an observer couples the EF to the component using an interface provided by the integration infrastructure. We refer to the DEVS model that consists of the observer and EF as a *test agent*.

Net-centric Service Oriented Architecture (SOA) provides a currently relevant technologically feasible realization of the concept. As discussed earlier, the DEVS/SOA infrastructure enables DEVS models, and test agents in particular, to be deployed to the network nodes of interest. As illustrated in Figure 23, in this incarnation, the network inputs sent by EF generators are SOAP messages sent to other EFs as destinations; transducers record the arrival of messages and extract the data in their fields, while acceptors decide on whether the gathered data indicates continuation or termination is in order [18,33].

Since EFs are implemented as DEVS models, distributed EFs are implemented as DEVS models, or agents as we have called them, residing on network nodes. Such a federation, illustrated in Figure 24, consists of DEVS simulators executing on web servers on the nodes exchanging messages and obeying time relationships under the rules contained within their hosted DEVS models. Complete analysis of the design problem and its mapping to the three linguistic levels is available at [5].

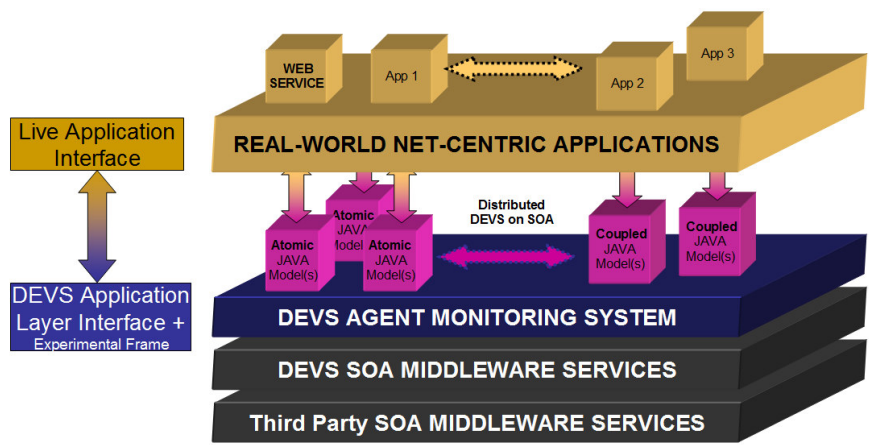

**Figure 23:** Deploying Experimental Frame Agents and Observers

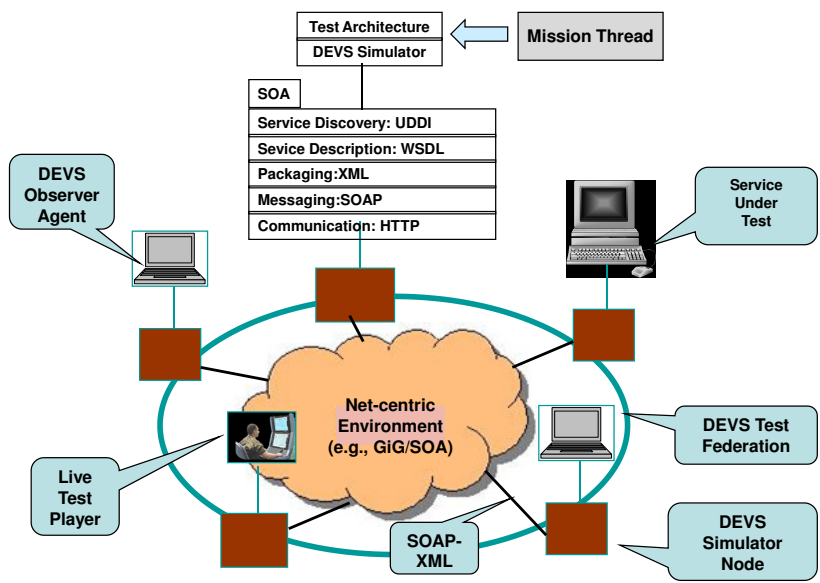

**Figure 24:** DEVS Test Federation in GIG/SOA Environment

# 11. Conclusions

Achieving interoperability is one of the chief SoS engineering objectives in the development of command and control (C2) capabilities for joint and coalition warfare. The importance of M&S in SoS design and evaluation cannot be underestimated. M&S can be used strategically to provide early feasibility studies and aid the design process. As components comprising SoS are designed and analyzed, their integration and communication is the most critical part that must be addressed by the employed SoS M&S framework. The integration infrastructure must support interoperability at syntactic, semantic and pragmatic levels to enable such integration.

Currently there are several other approaches to distributed simulation and to integration of M&S with advanced C2 systems. These approaches build on the internet or other net-centric middleware to provide component connectivity and simulation services [1,20]. The latter may also include HLA implementations; however, the extent of adoption of HLA in this context remains to be seen. The DEVS standard provides a formal systems-based abstraction that can support higher level interoperability, whether alone or on top of HLA. The DEVS/SOA implementation provides a SOA implementation independent of HLA and is a viable approach to M&S integration with C2 SoS in the weaker gateway form, and in the strong direct compliance form. Further, DEVS has been applied to frameworks like DoDAF, UML and other systems engineering frameworks like System Entity Structure (SES). Figure 25 illustrates how M&S is increasingly incorporated in C2 SoS as source of smart components as well as a methodology to deal with the interoperability problem. Indeed, DEVS components including decision making agents, sensor simulators, and environmental representations can bring the power of M&S to the development of C2 SoS as well as serving as support for command and control in real operation. The underlying SOA standard that facilitates this interoperation can be expected to be widely adopted (for example, it has been adopted by the DoD's Global Information Grid initiative.

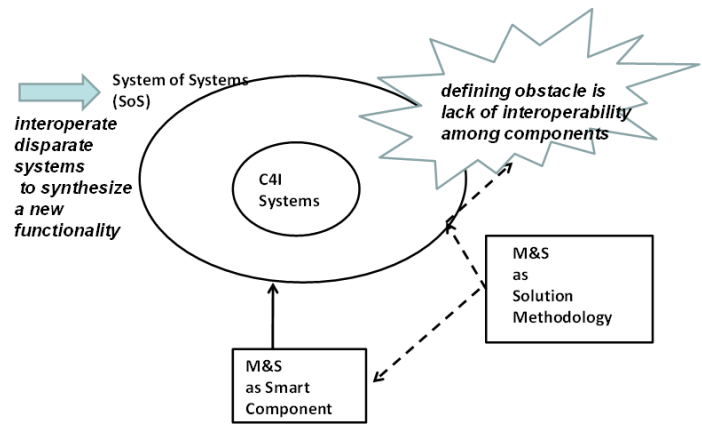

**Figure 25:** M&S as source of smart components in C4I systems

## Authors Biography

**Saurabh Mittal** is the CEO at DUNIP Technologies, India. Previously he worked as Research Assistant Professor at the Department of Electrical and Computer Engineering at the University of Arizona where he received his Ph. D in 2007. His areas of interest include Web-based M&S using SOA, interoperability, executable architectures, distributed simulation, and System of Systems engineering using DoDAF. He can be reached at saurabh.mittal@duniptechnologies.com

**Bernard P. Zeigler** is Professor of Electrical and Computer Engineering at the University of Arizona, Tucson and Director of the Arizona Center for Integrative Modeling and Simulation. He is developing DEVS-methodology approaches for testing mission thread end-to-end interoperability and combat effectiveness of Defense Department acquisitions and transitions to the Global Information Grid with its Service Oriented Architecture (GIG/SOA). He can be reached at zeigler@ece.arizona.edu

**José L. Risco-Martín** is an Assistant Professor in Complutense University of Madrid, Spain. He received his PhD from Complutense University of Madrid in 2004. His research interests are computational theory of modeling and simulation, with emphasis on DEVS, Dynamic memory management of embedded systems, and net-centric computing. He can be reached at jlrisco@dacya.ucm.es